\documentclass[prd,reprint,showpacs,showkeys]{revtex4-1}
\usepackage{amssymb}
\usepackage{amsmath}
\usepackage{graphicx}
\usepackage[font={footnotesize,it}]{caption}

\pdfoutput=1

\begin{document}

\title{A scan of f(R) models admitting Rindler type acceleration }
\author{S. Habib Mazharimousavi}
\email{habib.mazhari@emu.edu.tr}
\author{M. Kerachian}
\email{morteza.kerachian@cc.emu.tr}
\author{M. Halilsoy}
\email{mustafa.halilsoy@emu.edu.tr}
\date{\today }

\begin{abstract}
As a manifestation of large distance effect Grumiller modified Schwarzschild
metric with an extraneous term reminiscent of Rindler acceleration. Such a
term has the potential to explain the observed flat rotation curves in
general relativity. The same idea has been extended herein to the larger
arena of $f\left( R\right) $ theory. With particular emphasis on weak energy
conditions (WECs) for a fluid we present various classes of $f\left(
R\right) $ theories admitting a Rindler-type acceleration in the metric.
\end{abstract}

\maketitle
\affiliation{Physics Department, Eastern Mediterranean University, G. Magusa north
Cyprus, Mersin 10 Turkey}
\affiliation{Department of Physics, Eastern Mediterranean University, G. Magusa, north
Cyprus, Mersin 10 - Turkey}

\section{INTRODUCTION}

Flat rotation curves around galaxies constitute one of the most stunning
astrophysical findings since 1930s. The cases can simply be attributed to
the unobservable dark matter which still lacks a satisfactory candidate. On
the general relativity side which reigns in the large universe an
interesting approach is to develop appropriate models of constant
centrifugal force. One such attempt was formulated by Grumiller \cite{1,2}
in which the centrifugal force was given by $F=-\left( \frac{m}{r^{2}}%
+a\right) $. Here $m$ represents the mass (both normal and dark) while the
parameter "$a$" is a positive constant - called Rindler acceleration \cite{3}
- which gives rise to a constant attractive force. The Newtonian potential
involved herein is $\Phi \left( r\right) \sim -\frac{m}{r}+ar$ , so that for 
$r\rightarrow \infty $ the term $\Phi \left( r\right) \sim ar$ becomes
dominant. Since in Newtonian circular motion $F=\frac{mv^{2}}{r}$, for a
mass $m,$ tangential speed $v\left( r\right) $ and radius $r$ are related by 
$v\left( r\right) \sim r^{\frac{1}{2}}$ for large $r$, overall which amounts
slightly nearer to the concept of flat rotation curves. No doubts, the
details and exact flat rotation curves must be much more complicated than
the toy model depicted here. Physically the parameter "$a$" becomes
meaningful when one refers to an accelerated frame in a flat space, known as
Rindler frame and accordingly the terminology Rindler acceleration is
adopted.

In \cite{4} impact of a Rindler-type acceleration is studied on the Oort
Cloud and in \cite{5} the solar system constraints on Rindler acceleration
is investigated while in \cite{6} bending of light in the model of gravity
at large distances proposed by Grumiller \cite{1,2} is considered.

Let us add also that to tackle the flat rotation curves, Modified Newtonian
Dynamics (MOND) in space was proposed \cite{7}. Assuming a physical source
to the Rindler acceleration term in the spacetime metric has been
challenging in recent years. Anisotropic fluid field was considered
originally by Grumiller \cite{1,2}, whereas nonlinear electromagnetism was
proposed as an alternative source \cite{8}. A fluid model with
energy-momentum tensor of the form $T_{\mu }^{\nu }=diag[-\rho ,p,q,q]$ was
proposed recently in the popular $f\left( R\right) $ gravity \cite{9}. For a
review of the latter we propose \cite{10,11,12}. By a similar strategy we
wish to employ the vast richness of $f(R)$ gravity models to identity
possible candidates that may admit Rindler type acceleration. Our approach
in this study beside the Rindler acceleration is to elaborate on the energy
conditions in $f\left( R\right) $ gravity. Although violation of the energy
conditions is not necessarily a problem (for instance, any quantum field
theory violates all energy conditions) but it is still interesting to
investigate the non-violation of the energy conditions. Note that energy
conditions within the context of dark matter in $f(R)$ gravity was
considered by various authors \cite{13}. This at least will filter the
viable models that satisfy the energy conditions. In brief, for our choice
of energy-momentum the weak energy conditions (WECs) can be stated as
follow: i) WEC1 says that energy density $\rho \geqslant 0$. ii) WEC2, says
that $\rho +p\geqslant 0$, and iii) WEC3 states that $\rho +q\geqslant 0$.
The more stringent energy conditions, the strong energy conditions (SECs)
amounts further to $\rho +p+2q\geqslant 0$, which will not be our concern in
this paper. However, some of our models satisfy SECs as well. Our technical
method can be summarized as follows. Upon obtaining $\rho ,$ $p$ and $q$ as
functions of $r$ we shall search numerically for the geometrical regions in
which the WECs are satisfied. (A detailed work on energy condition in $f(R)$
gravity was done by J. Santos et al in \cite{14}).

From the outset our strategy is to assume validity of the Rindler modified
Schwarzschild metric a priori and search for the types of $f(R)$ models
which are capable to yield such a metric. Overall we test ten different
models of $f\left( R\right) $ gravity models and observe that in most cases
it is possible to tune free parameters in rendering the WECs satisfied. In
doing this we entirely rely on numerical plots and we admit that our list is
not an exhaustive one in $f\left( R\right) $ arena.

Organization of the paper goes as follows. Sec. II introduces the formalism
with derivation of density and pressure components. Sec. III presents eleven
types of $f\left( R\right) $ models relevant to the Mannheim's metric. The
paper ends with Conclusion in Sec. IV.

\section{The Formalism}

\begin{figure}[tbp]
\includegraphics[width=80mm,scale=0.7]{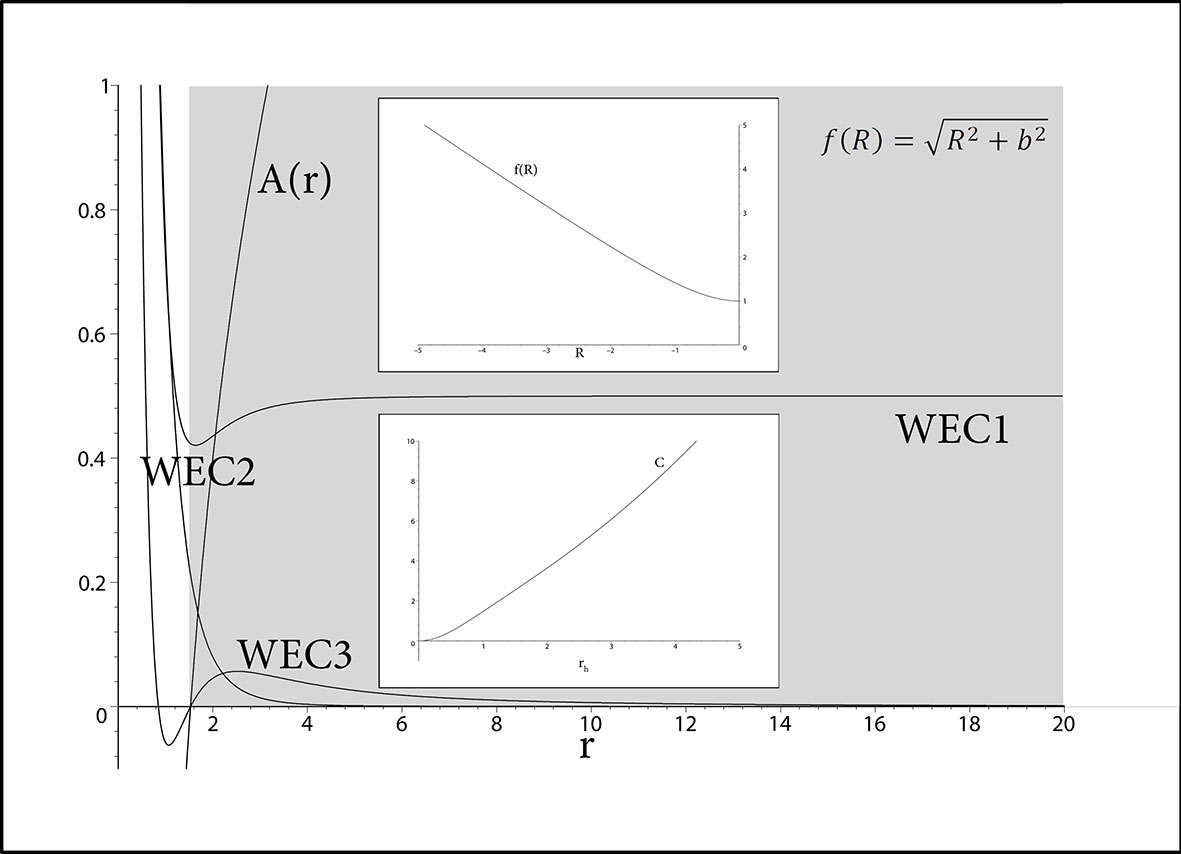}
\caption{A plot of $WEC1,$ $WEC2$ and $WEC3$ for $m=1$, $a=0.1$ and $b=1.$
To have an idea of the range in which WECs are satisfied we also plot the
metric function which identifies the location of the horizon. It is observed
from the figure that WECs are all satisfied for $r\geq r_{h}$ in which $%
r_{h} $ is the event horizon of the Grumiller's metric. Since $R<0$ the plot
of $f(R)$ is from $-\infty $ up to zero and as can be seen $\frac{df}{dR}<0$
while $\frac{d^{2}f}{dR^{2}}>0$. We also plot the heat capacity $C$ w.r.t
the horizon radius $r_{h}$.}
\end{figure}

\begin{figure}[tbp]
\includegraphics[width=80mm,scale=0.7]{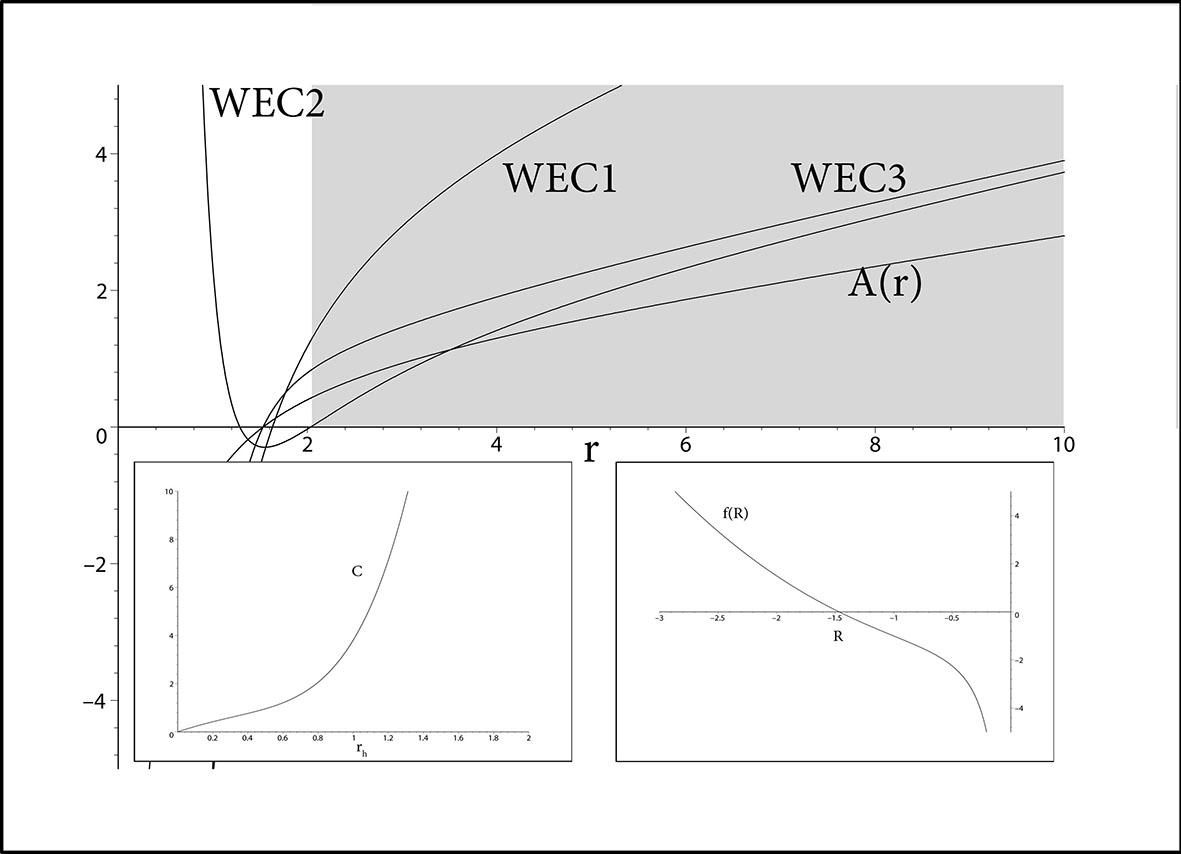}
\caption{Our choice of the parameters are $\protect\nu =1,$ $\protect\mu =2,$
$b=1,$ $c=-1,$ $\Lambda _{1}=0=\Lambda _{2}$. WECs are shown to satisfy
while stability is valid only for $R<-1$. This can easily be checked from $%
f\left( R\right) =R+\frac{1}{R}+R^{2}.$ Thermodynamic stability (i.e. $C>0$)
is also shown in the inscription.}
\end{figure}
Let's start with the following action ($\kappa =8\pi G=1$) 
\begin{equation}
S=\frac{1}{2}\int \sqrt{-g}f\left( R\right) d^{4}x+S_{M}
\end{equation}%
where $f\left( R\right) $ is a function of the Ricci scalar $R$ and $S_{M}$
is the physical source for a perfect fluid-type energy momentum 
\begin{equation}
T_{\mu }^{\nu }=\left( 
\begin{array}{cccc}
-\rho & 0 & 0 & 0 \\ 
0 & p & 0 & 0 \\ 
0 & 0 & q & 0 \\ 
0 & 0 & 0 & q%
\end{array}%
\right)
\end{equation}%
We adopt the static spherically symmetric line element 
\begin{equation}
ds^{2}=-A\left( r\right) dt^{2}+\frac{1}{A(r)}dr^{2}+r^{2}\left( d\theta
^{2}+\sin ^{2}\theta d\varphi ^{2}\right)
\end{equation}%
with 
\begin{equation}
A\left( r\right) =1-\frac{2m}{r}+2ar
\end{equation}%
which will be referred henceforth as the Mannheim's metric \cite{15} (Note
that it has been rediscovered by Grumiller in \cite{1,2}). Einstein's field
equations follow the variation of the action with respect to $g_{\mu \nu }$
which reads as%
\begin{equation}
G_{\mu }^{\nu }=\frac{1}{F}T_{\mu }^{\nu }+\check{T}_{\mu }^{\nu }
\end{equation}%
in which $G_{\mu }^{\nu }$ is the Einstein's tensor. The share of the
curvature in the energy-momentum is given by%
\begin{equation}
\check{T}_{\mu }^{\nu }=\frac{1}{F}\left[ \nabla ^{\nu }\nabla _{\mu
}F-\left( \square F-\frac{1}{2}f+\frac{1}{2}RF\right) \delta _{\mu }^{\nu }%
\right]
\end{equation}%
while $T_{\mu }^{\nu }$ refers to the fluid source \cite{1,2}. Following the
standard notation, $\square =\nabla ^{\mu }\nabla _{\mu }=\frac{1}{\sqrt{-g}}%
\partial _{\mu }\left( \sqrt{-g}\partial ^{\mu }\right) $ and $\nabla ^{\nu
}\nabla _{\mu }u=g^{\lambda \nu }\nabla _{\lambda }u_{,\mu }=g^{\lambda \nu
}\left( \partial _{\lambda }u_{,\mu }-\Gamma _{\lambda \mu }^{\beta
}u_{,\beta }\right) $ for a scalar function $u$. The three independent
Einstein's field equations are explicitly given by 
\begin{equation}
FR_{t}^{t}-\frac{f}{2}+\square F=\nabla ^{t}\nabla _{t}F+T_{t}^{t}
\end{equation}%
\begin{equation}
FR_{r}^{r}-\frac{f}{2}+\square F=\nabla ^{r}\nabla _{r}F+T_{r}^{r}
\end{equation}%
\begin{eqnarray}
FR_{\theta _{i}}^{\theta _{i}}-\frac{f}{2}+\square F &=&\nabla ^{\theta
_{i}}\nabla _{\theta _{i}}F+T_{\theta _{i}}^{\theta _{i}} \\
&&\left( F=\frac{df}{dR}\right) ,
\end{eqnarray}%
in which $\theta _{i}=\left( \theta ,\varphi \right) .$ Adding these
equations (i.e., $tt$, $rr$, $\theta \theta $ and $\varphi \varphi $) one
gets the trace equation 
\begin{equation}
FR-2f+3\square F=T
\end{equation}%
which is not an independent equation. Using the field equations one finds%
\begin{equation}
\rho =\nabla ^{t}\nabla _{t}F-FR_{t}^{t}+\frac{f}{2}-\square F,
\end{equation}%
\begin{equation}
p=-\nabla ^{r}\nabla _{r}F+FR_{t}^{t}-\frac{f}{2}+\square F,
\end{equation}%
and%
\begin{equation}
q=-\nabla ^{\theta }\nabla _{\theta }F+FR_{\theta }^{\theta }-\frac{f}{2}%
+\square F.
\end{equation}%
In what follows we find the energy momentum components for different models
of $f(R)$ gravity together with their thermodynamical properties.

\section{$f(R)$ Models apt for the Rindler Acceleration}

\begin{figure}[tbp]
\includegraphics[width=80mm,scale=0.7]{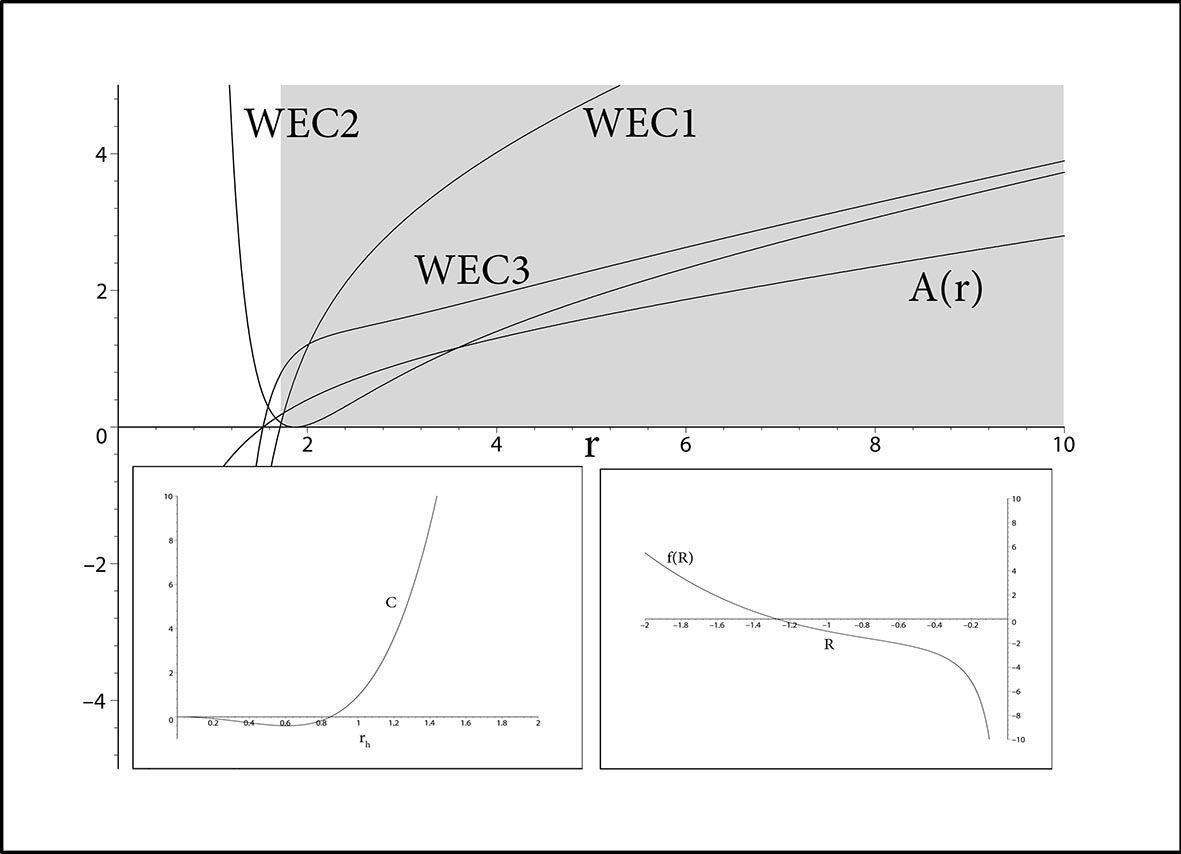}
\caption{ Our parameters in this case are $\protect\nu =1,$ $\protect\mu =2,$
$b=-1,$ $c=-1$ with $\Lambda _{1}=0=\Lambda _{2}$. So that $f(R)$ takes the
form $f\left( R\right) =R+\frac{1}{R}-R^{3}$ which satisfies the WECs. This
choice yields a stable model for $R<-\frac{1}{\sqrt[4]{3}}.$ Beyond certain
horizon radius the specific function $C$ is also positive.}
\end{figure}

In this section we investigate a set of possible $f(R)$ gravity models which
admit the line element (3) as the static spherically symmetric solution of
its field equations. Then by employing Eq.s (12) to (14) we shall find the
energy density $\rho $ and the pressures $p$ and $q.$ Having found $\rho $, $%
p$ and $q$ we investigate the energy conditions together with the
feasibility of the $f(R)$ models numerically. More precisely we work on weak
energy conditions which includes three individual conditions 
\begin{equation}
WEC1=\rho \geq 0,
\end{equation}%
\begin{equation}
WEC2=\rho +q\geq 0
\end{equation}%
and 
\begin{equation}
WEC3=\rho +p\geq 0.
\end{equation}%
In the numerical plotting, we plot explicitly $WEC1,$ $WEC2$ and $WEC3$ in
terms of $r$ to work out the region(s) in which the WECs are satisfied. In
addition to WECs we plot $f\left( R\right) $ in terms of $R$ to find out the
physically acceptable model by imposing the well known conditions on $%
f\left( R\right) $ which are given by%
\begin{equation}
F\left( R\right) =\frac{df\left( R\right) }{dR}>0
\end{equation}%
for not to have ghost field and%
\begin{equation}
\frac{d^{2}f\left( R\right) }{dR^{2}}>0
\end{equation}%
to have a stable model. Before we start to study the $f(R)$ models, we add
that in the case of Mannheim's metric the Ricci scalar is given by $R=-\frac{%
12a}{r}$ which is negative ($a>0$).

\subsection{The Models}

\begin{figure}[tbp]
\includegraphics[width=80mm,scale=0.7]{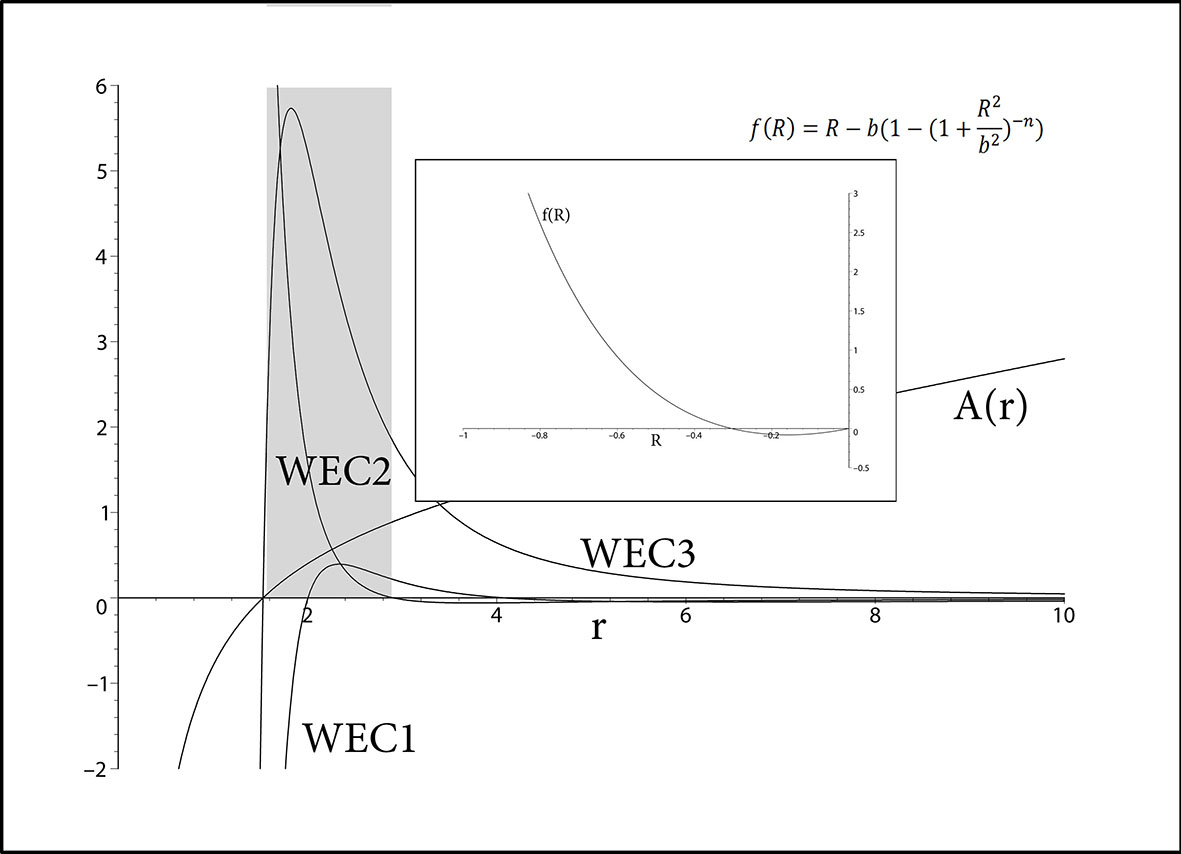}
\caption{From Eq. (31) we choose the parameters as $\protect\mu =1,$ $b=1$
and $n=-3$. We find a restricted domain in which WECs are satisfied. From
those parameters beside WECs from $\frac{d^{2}f}{dR^{2}}=6\left(
1+R^{2}\right) \left( 1+5R^{2}\right) >0$ the stability condition also is
satisfied.}
\end{figure}

\begin{figure}[tbp]
\includegraphics[width=80mm,scale=0.7]{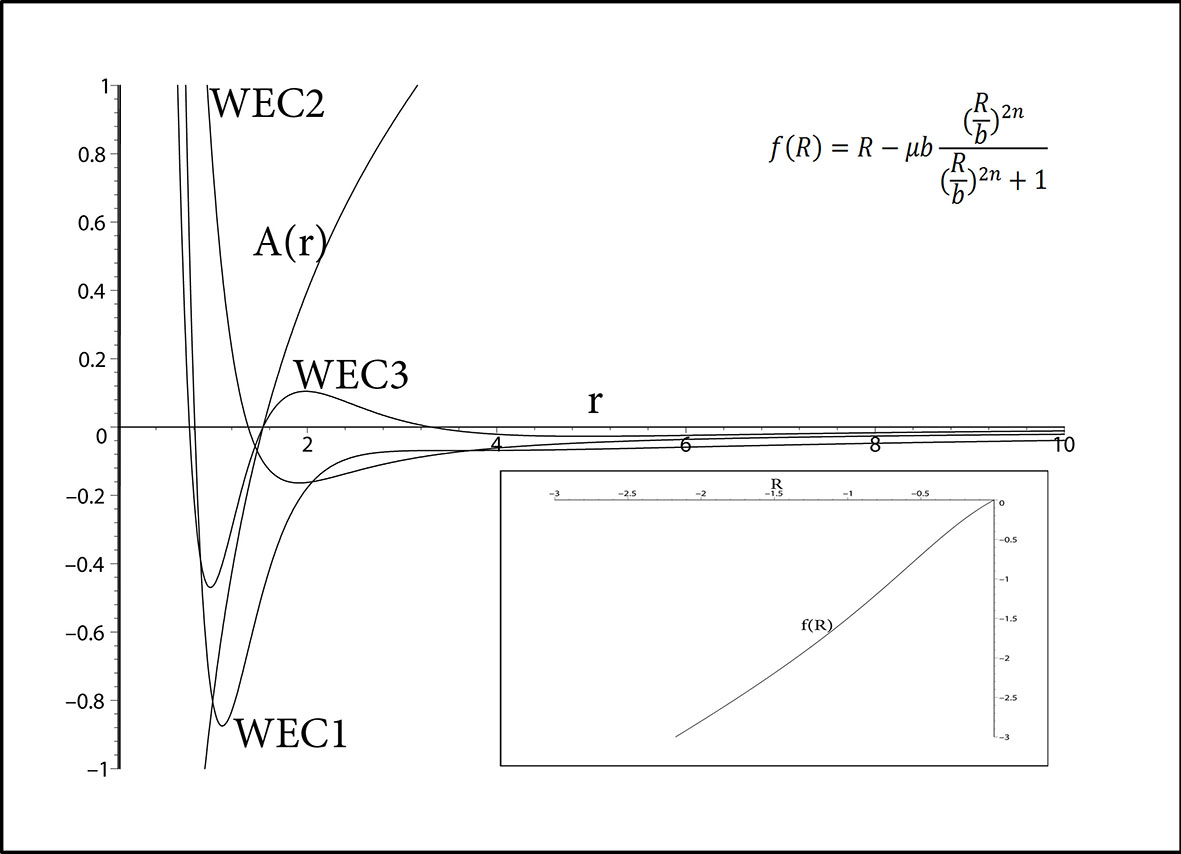}
\caption{In this model given by the $f(R)$ in Eq. (32) we have not been able
to find a physically admissible region to satisfy WECs.}
\end{figure}

\begin{figure}[tbp]
\includegraphics[width=80mm,scale=0.7]{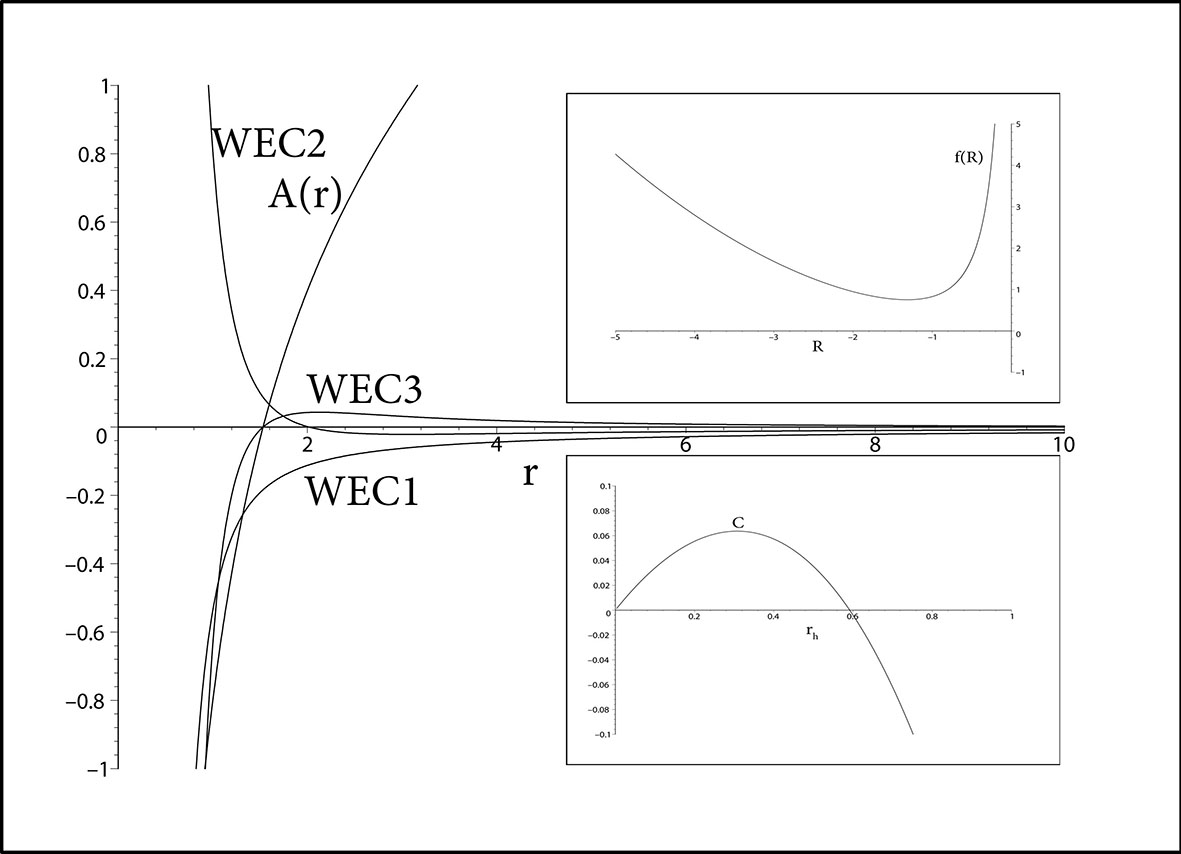}
\caption{From $f\left( R\right) $ model in Eq. (33) the choice $c=1/3,$ $%
\protect\varepsilon =1$ and $b=1$ , we observe that WECs are not satisfied.
Specific heat function is also pictured in the inscription.}
\end{figure}

\begin{figure}[tbp]
\includegraphics[width=80mm,scale=0.7]{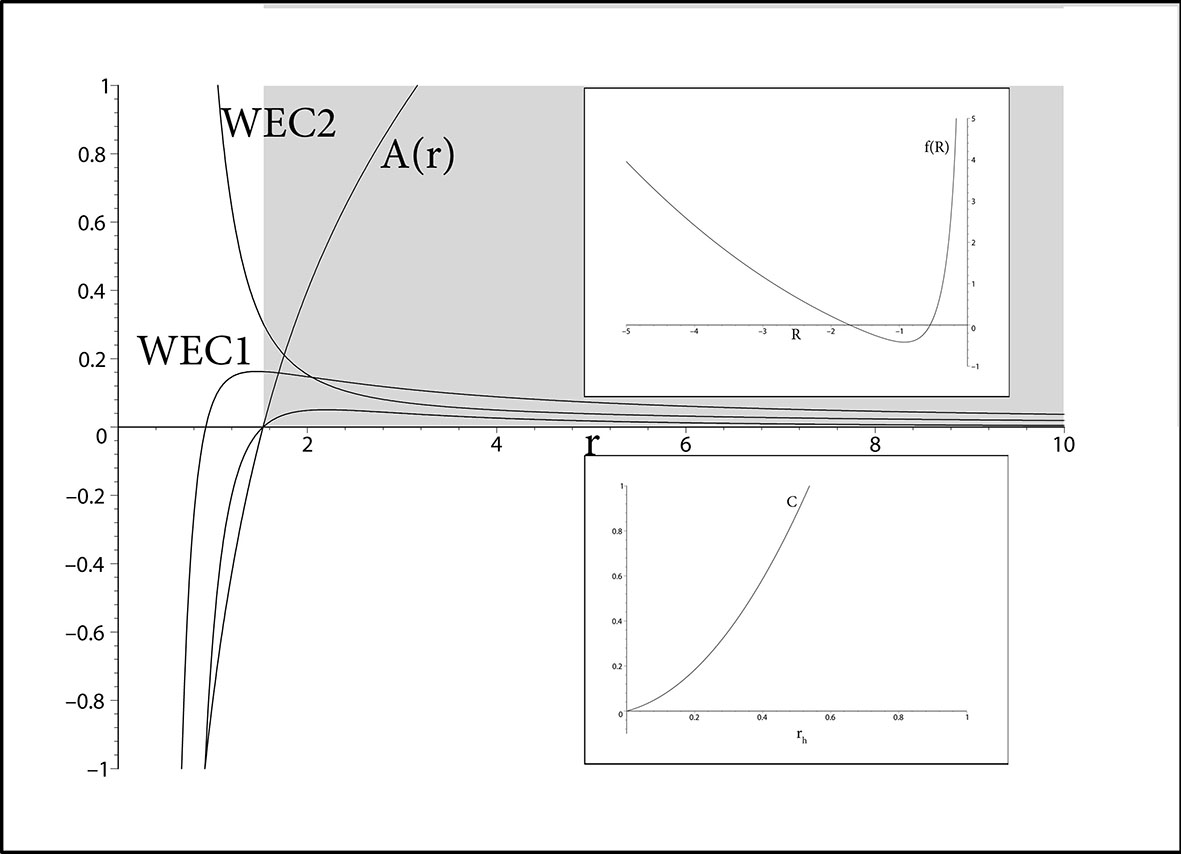}
\caption{The choice of parameters $c=1.1,$ $\protect\varepsilon =1$ and $b=1$
in Eq. (33) yields a region where WECs are satisfied. It can be checked that 
$\frac{d^{2}f}{dR^{2}}>0$ is also satisfied. For $\left\vert R\right\vert
>\left\vert R_{0}\right\vert $ where $f^{\prime }\left( R_{0}\right) =0,$ $%
\frac{df}{dR}>0$ which implies ghost free solution. Everywhere positive
specific heat $C$ is also shown in the inscription.}
\end{figure}

\textbf{1)} Our first model which we find interesting is given by \cite{16} 
\begin{equation}
f\left( R\right) =\sqrt{R^{2}+b^{2}}
\end{equation}%
for $b=$constant. For $\left\vert R\right\vert \gg b,$ this model is a good
approximation to Einstein's $f(R)=R$ gravity. For the other extent, namely $%
\left\vert R\right\vert \ll b$, $b$ may be considered as a cosmological
constant. Having this $f\left( R\right) $ one finds 
\begin{equation}
\frac{df}{dR}=\frac{R}{\sqrt{R^{2}+b^{2}}}
\end{equation}%
\begin{equation}
\frac{d^{2}f}{dR^{2}}=\frac{b^{2}}{\left( R^{2}+b^{2}\right) ^{3/2}}
\end{equation}%
which are positive functions with respect to $R$. This means that this model
of $f(R)$ gravity is satisfying the necessary conditions to be physical. Yet
we have to check the WECs at least to see whether it can be a good candidate
for a spacetime with Rindler acceleration, namely the Mannheim's metric.
Figure 1 displays $WEC1,$ $WEC2$ and $WEC3$ together with part of $A(r)$ in
terms of $r.$ We see that the WECs are satisfied right after the horizon.
Therefore this model can be a good candidate for what we are looking for.
This model is also interesting in other aspects. For instance in the limit
when $b$ is small one may write 
\begin{equation}
f\left( R\right) \simeq \left\vert R\right\vert +\frac{b^{2}}{2}\frac{%
\left\vert R\right\vert }{R^{2}}
\end{equation}%
which is a kind of small fluctuation from $R$ gravity for $\left\vert
R\right\vert \gg b$.

In particular, this model of $f(R)$ gravity is satisfying all necessary
conditions to be a physical model to host Mannheim's metric. Hence we go one
step more to check the heat capacity of the spacetime to investigate if the
solution is stable from the thermodynamical point of view. To do so, first
we find the Hawking temperature 
\begin{equation}
T_{H}=\left. \frac{\frac{\partial }{\partial r}g_{tt}}{4\pi }\right\vert
_{r=r_{h}}=\frac{m+ar_{h}^{2}}{2\pi r_{h}^{2}}.
\end{equation}%
Then, from the general form of the entropy in $f(R)$ gravity we find%
\begin{equation}
S=\left. \frac{\mathcal{A}}{4G}F\right\vert _{r=r_{h}}=\pi r_{h}^{2}F_{h}
\end{equation}%
in which $\left. \mathcal{A}\right\vert _{r=r_{h}}=4\pi r_{h}^{2}$ is the
surface area of the black hole at the horizon and $\left. F\right\vert
_{r=r_{h}}=\frac{-12a}{\sqrt{\frac{144a^{2}}{r_{h}^{2}}+b^{2}r_{h}}}$.
Having $T_{H}$ and $S$ available one may find the heat capacity of the black
hole as%
\begin{equation}
C=T\left( \frac{\partial S}{\partial T}\right) =\frac{12\left(
1+4ar_{h}\right) \left( 288a^{2}+b^{2}r_{h}^{2}\right) r_{h}^{2}\pi a}{%
\left( 144a^{2}+b^{2}r_{h}^{2}\right) ^{3/2}}.
\end{equation}%
We comment here that $C$ is always positive and nonsingular irrespective of
the values of the free parameters given the fact that $a>0$. This indeed
means that the black hole solution will not undergo a phase change as
expected form a stable physical solution.

\textbf{2)} The second model which we shall study, in this part, has been
introduced and studied by Nojiri and Odintsov in \cite{17}. As they have
reported in their paper \cite{17}, "\textit{this model naturally unifies two
expansion phases of the Universe: in-flation at early times and cosmic
acceleration at the current epoch". }This model of $f(R)$ is given by 
\begin{equation}
f(R)=R-\frac{c}{\left( R-\Lambda _{1}\right) ^{\nu }}+b\left( R-\Lambda
_{2}\right) ^{\mu }
\end{equation}%
in which $b,$ $c,$ $\Lambda _{1},$ $\Lambda _{2},$ $\mu $ and $\nu $ are
some adjustable parameters. Our plotting strategy of each model is such that
if the WECs are violated (note that such cases are copious) we ignore such
figures and regions satisfying WECs are shaded. The other conditions $\frac{%
df}{dR}>0,$ $\frac{d^{2}f}{dR^{2}}>0$ are satisfied in some cases whereas in
the others not. In Figs. 2 and 3 we plot $WEC1,$ $WEC2$ and $WEC3$ in terms
of $r$ for specific values of $\nu ,$ $\mu ,b,$ $c$ i.e. in Fig. 2 $\nu
=1,\mu =2,$ $b=1,$ $c=-1$, $\Lambda _{1}=0$, $\Lambda _{2}=0.$ In Fig. 3 $%
\nu =1,$ $\mu =3,$ $b=-1,$ $c=-1,\Lambda _{1}=0$, $\Lambda _{2}=0$.

Among the particular cases which are considered here, one observes that Fig.
2 and Fig. 3 which correspond to 
\begin{equation}
f\left( R\right) =R+\frac{1}{R}+R^{2}
\end{equation}%
and%
\begin{equation}
f\left( R\right) =R+\frac{1}{R}-R^{3}
\end{equation}%
respectively, are physically acceptable as far as WECs are concerned. We
also note that in these two figures we plot the heat capacity in terms of $%
r_{h}$ to show whether thermodynamically the solutions are stable. $\frac{%
d^{2}f}{dR^{2}}$ reveals that (28) and (29) are locally stable.

\textbf{3)} Our next model is a Born-Infeld type gravity which has been
studied in a more general form of Dirac-Born-Infeld modified gravity by
Quiros and Ure\~{n}a-L\'{o}pez in \cite{18}. The Born-Infeld model of
gravity is given by $f\left( R\right) =2b\left( 1-\sqrt{1+\frac{\left\vert
R\right\vert }{b}}\right) ,$ which implies

\begin{equation}
F\left( R\right) =\frac{1}{\sqrt{1+\frac{\left\vert R\right\vert }{b}}} 
\notag
\end{equation}%
and 
\begin{equation}
\frac{d^{2}f}{dR^{2}}=\frac{1}{2\left( 1+\frac{\left\vert R\right\vert }{b}%
\right) ^{3/2}}
\end{equation}%
Clearly both are positive functions of $R$ therefore the solution given in
this model is stable and ghost free. In spite of that, the WECs are not
satisfied therefore this model is not a proper model for Mannheim's metric
as far as the energy conditions are concerned.

\begin{figure}[tbp]
\includegraphics[width=80mm,scale=0.7]{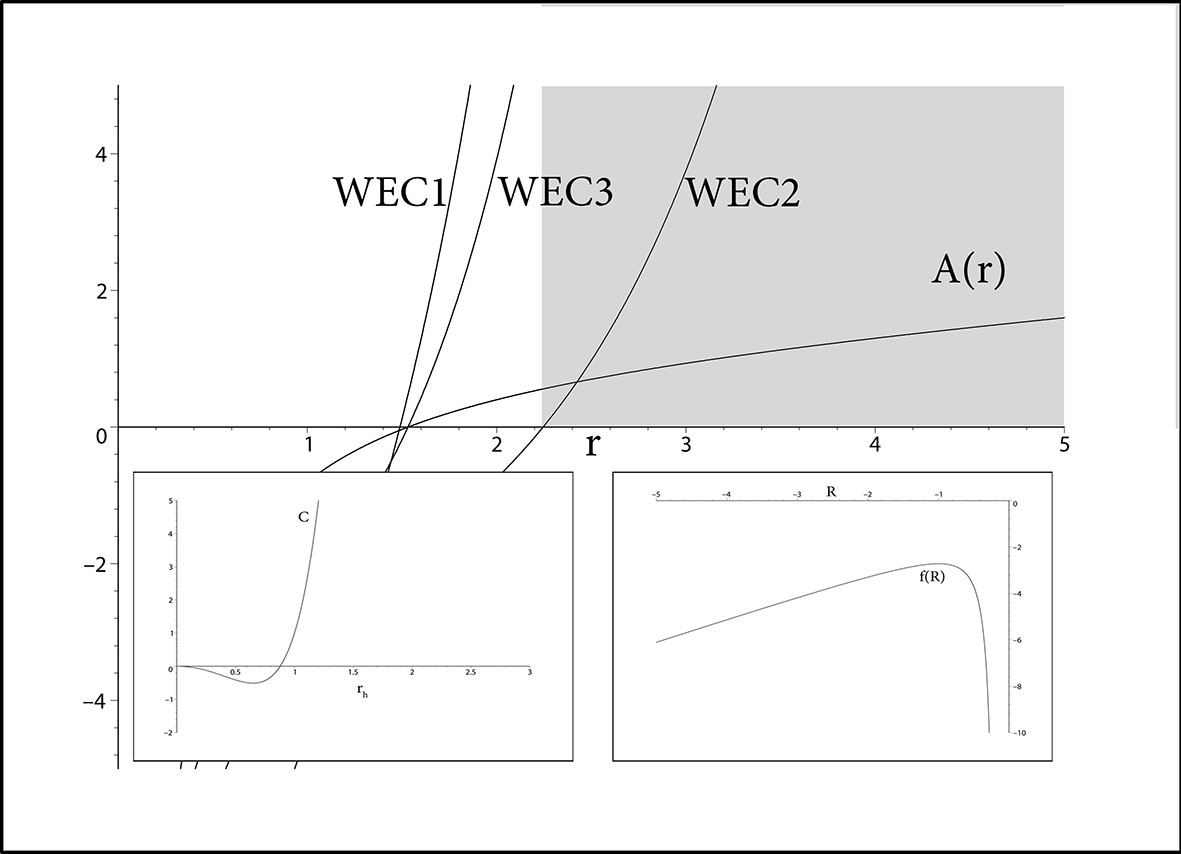}
\caption{The model with $f\left( R\right) =R$e$^{-\frac{1}{R}}$ gives a
region in which WECs are satisfied. Furthermore, since $\frac{d^{2}f}{dR^{2}}%
=\frac{1}{R^{3}}e^{-\frac{1}{R}}<0$, it gives an unstable model. Beyond
certain radius the specific heat is also positive which is required for
thermodynamical stability.}
\end{figure}

\begin{figure}[tbp]
\includegraphics[width=80mm,scale=0.7]{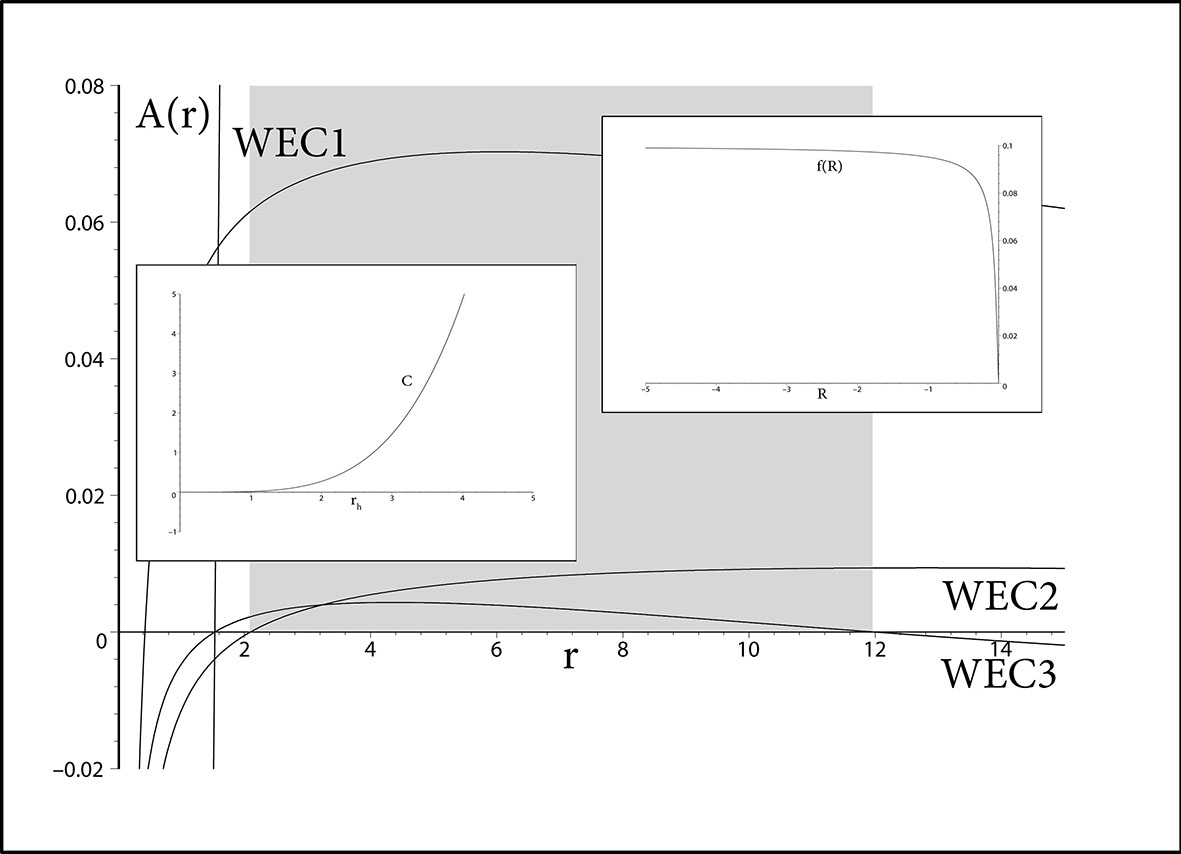}
\caption{Our model in this case is given by $f\left( R\right) =R\left( e^{%
\frac{b}{R}}-1\right) $ with $b=const$. With the choice $b=1.1$ its observed
that WECs are satisfied while the stability condition is violated in spite
of the fact that the specific heat $C$ is everywhere positive.}
\end{figure}

\textbf{4)} Another interesting model of $f(R)$ gravity is given by \cite{19}
\begin{equation}
f\left( R\right) =R-\mu b\left[ 1-\left( 1+\frac{R^{2}}{b^{2}}\right) ^{-n}%
\right]
\end{equation}%
in which $\mu ,b$ and $n$ are constants. Figure 4 with $\mu =1,b=1,n=-3$
shows that between horizon and a maximum radius we may have physical region
in which $f^{\prime \prime }>0$. Now let's consider \cite{20} the model 
\begin{equation}
f\left( R\right) =R-\mu b\frac{\left( \frac{R}{b}\right) ^{2n}}{\left( \frac{%
R}{b}\right) ^{2n}+1}
\end{equation}%
which amounts to the Fig. 5 and clearly there is no physical region.

\textbf{5)} Here, we use another model introduced in \cite{21} which is
given by 
\begin{equation}
f\left( R\right) =R\left( 1-c\right) +c\varepsilon \ln \left( \frac{\cosh
\left( \frac{R}{\varepsilon }-b\right) }{\cosh \left( b\right) }\right) +%
\frac{R^{2}}{6m^{2}}
\end{equation}%
in which $c,\varepsilon ,b$ and $\mu $ are all constants. Our analysis
yields to the Fig. 6 with $c=\frac{1}{3}$ and Fig. 7 with $c=1.1$. One
observes that although in Fig. 6 there is no physical region possible for
different $c$ in Fig. 7 and for $r>r_{h}$ our physical conditions are
satisfied provided $\left\vert R\right\vert <\left\vert R_{0}\right\vert $
where $R_{0}$ is the point for which $F\left( R\right) =0.$

\textbf{6)} In Ref. \cite{22} an exponential form of $f(R)$ is introduced
which is given by 
\begin{equation}
f\left( R\right) =Re^{\frac{b}{R}}
\end{equation}%
in which $b=$constant with its first derivative

\begin{equation}
F\left( R\right) =e^{\frac{b}{R}}\left( 1-\frac{b}{R}\right) .
\end{equation}%
Our numerical plotting admits the Fig. 8 for this model with $b=-1$. We
comment here that although the case $b=-1$ provides the WECs satisfied but
in both cases $f^{\prime \prime }\left( R\right) $ is negative which makes
the model not physical.

\bigskip \textbf{7)} Another exponential model which is also given in \cite%
{22} reads%
\begin{equation}
f(R)=Re^{bR},
\end{equation}%
in which $b=$constant and 
\begin{equation}
F(R)=e^{bR}\left( 1+bR\right) .
\end{equation}%
This does not satisfy the energy conditions and therefore it is not a
physically interesting case.

\textbf{8)} In Ref. \cite{23} a modified version of our models 6 and 7 is
given in which%
\begin{equation*}
f(R)=R\left( e^{\frac{b}{R}}-1\right)
\end{equation*}%
\ with $b=$constant and 
\begin{equation*}
F(R)=e^{\frac{b}{R}}\left( 1-\frac{b}{R}\right) -1.
\end{equation*}%
Figure 9 is our numerical results with $b=0.1$. For a bounded region from
above and from below the WECs are satisfied while $f^{\prime \prime }\left(
R\right) $ is negative which makes our model non-physical.

\textbf{9)} Among the exponential models of gravity let's consider \cite{24} 
\begin{equation}
f\left( R\right) =R+be^{\alpha R}
\end{equation}%
where $\alpha $ and $b$ are constants and 
\begin{equation}
F\left( R\right) =1+b\alpha \text{ }e^{\alpha R}.  \notag
\end{equation}%
Figure 17 displays our numerical calculations for specific values of $\alpha
=-1$ . Evidently from these figures we can conclude that this model is not a
feasible model.

\textbf{10) }Finally we consider a model of gravity given in Ref. \cite{25} 
\begin{equation}
f\left( R\right) =\left( \left\vert R\right\vert ^{b}-\Lambda \right) ^{%
\frac{1}{b}}
\end{equation}%
in which $b$ is a constant. The first derivative of the model is given by%
\begin{equation*}
F(R)=\left\vert R\right\vert ^{b-1}\left( \left\vert R\right\vert
^{b}-\Lambda \right) ^{\frac{1}{b}-1}.
\end{equation*}%
Figures 11 and 12 are with $b=\frac{1}{2}$ and $b=2,$ respectively, for $%
\Lambda =1.$ We observe that WECs are satisfied in a restricted region while
for $b=2$ / $\frac{1}{2}$ it gives a stable / unstable model.

\section{CONCLUSION}

\begin{figure}[tbp]
\includegraphics[width=80mm,scale=0.7]{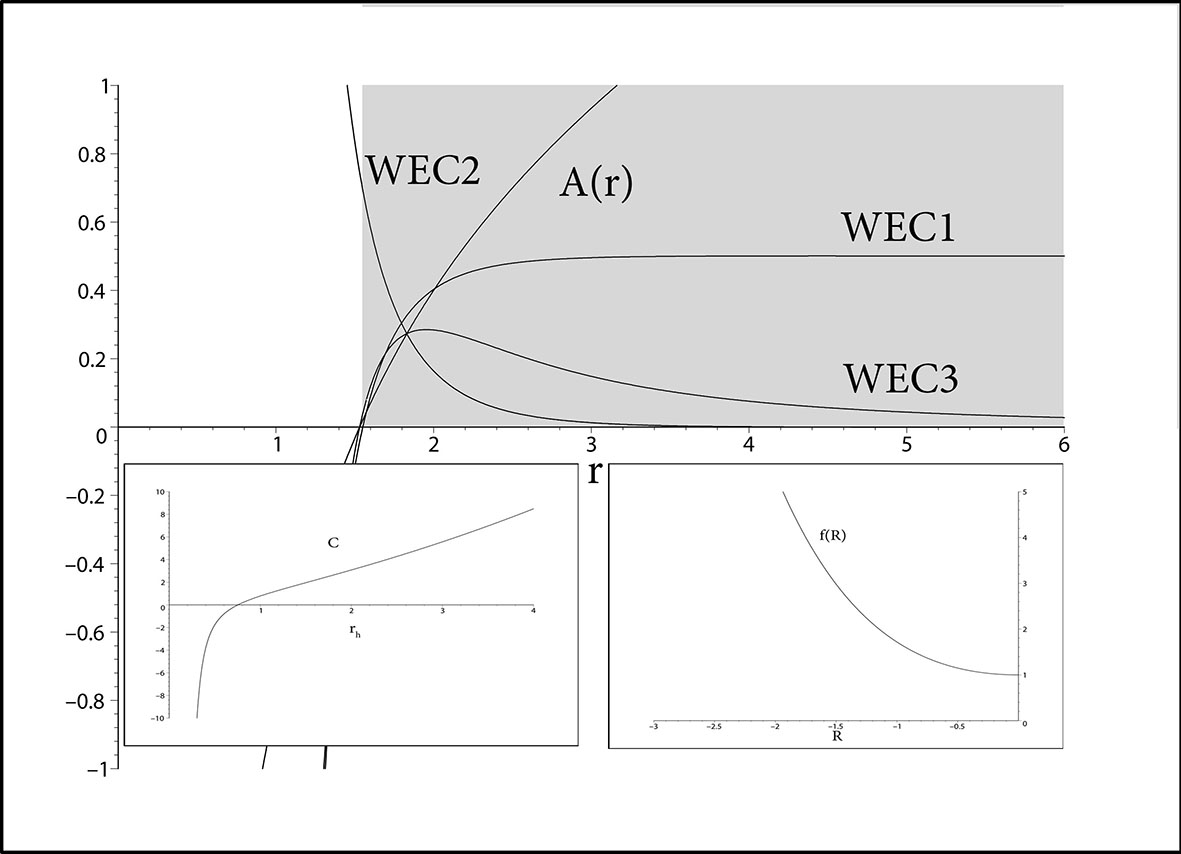}
\caption{In this model we use $f\left( R\right) =R+be^{\protect\alpha R}$
where $\protect\alpha $ and $b$ are constants. For $\protect\alpha =-1$ and $%
b=1,$ WECs are satisfied and $\frac{d^{2}f}{dR^{2}}>0$. Specific heat is
shown also to be positive.}
\end{figure}
\begin{figure}[tbp]
\includegraphics[width=80mm,scale=0.7]{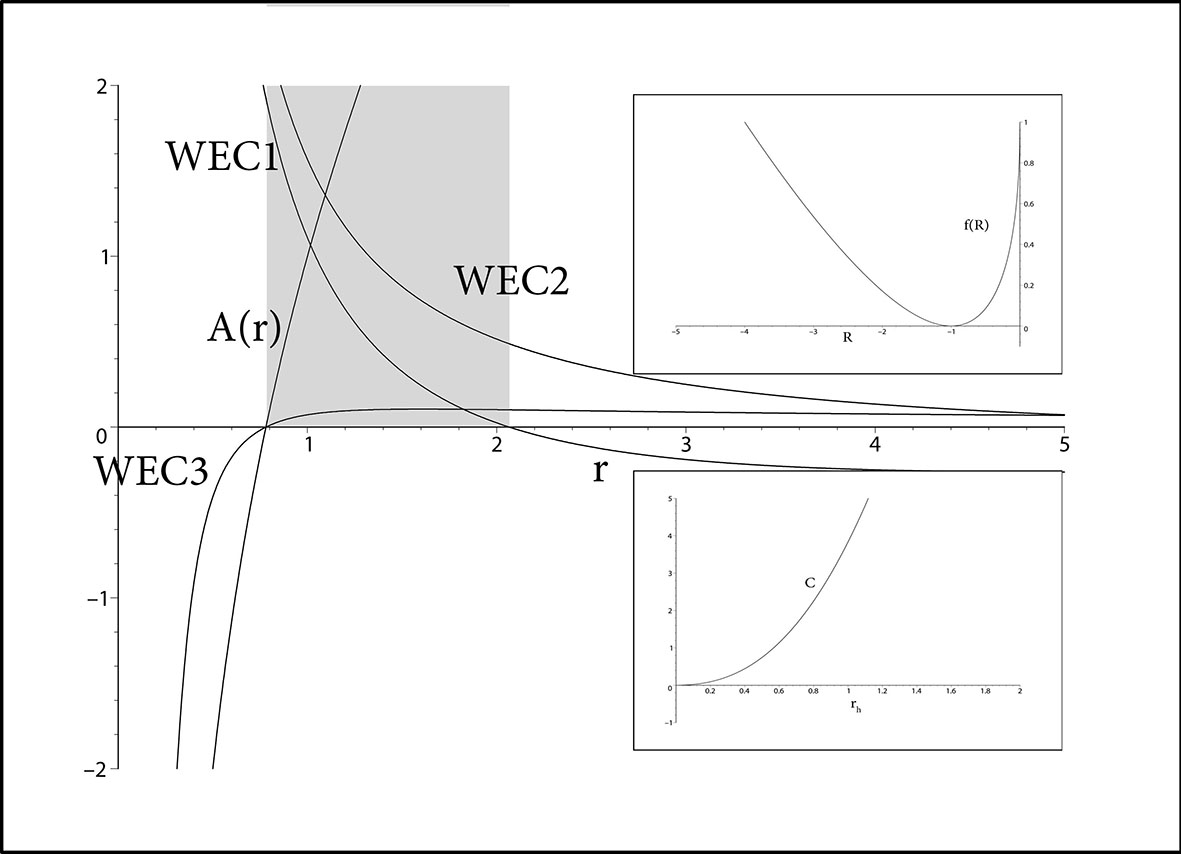}
\caption{Our model is given by $f\left( R\right) =\left( \mid R\mid
^{2}-1\right) ^{2}$ which has WECs satisfied but $\frac{d^{2}f}{dR^{2}}>0$,
for $\left\vert R\right\vert >\left\vert R_{0}\right\vert $ where $f^{\prime
}\left( R_{0}\right) =0.$ This indicates stability of the solution.
Furthermore the specific heat suggests a thermodynamically stable model too.}
\end{figure}
\begin{figure}[tbp]
\includegraphics[width=80mm,scale=0.7]{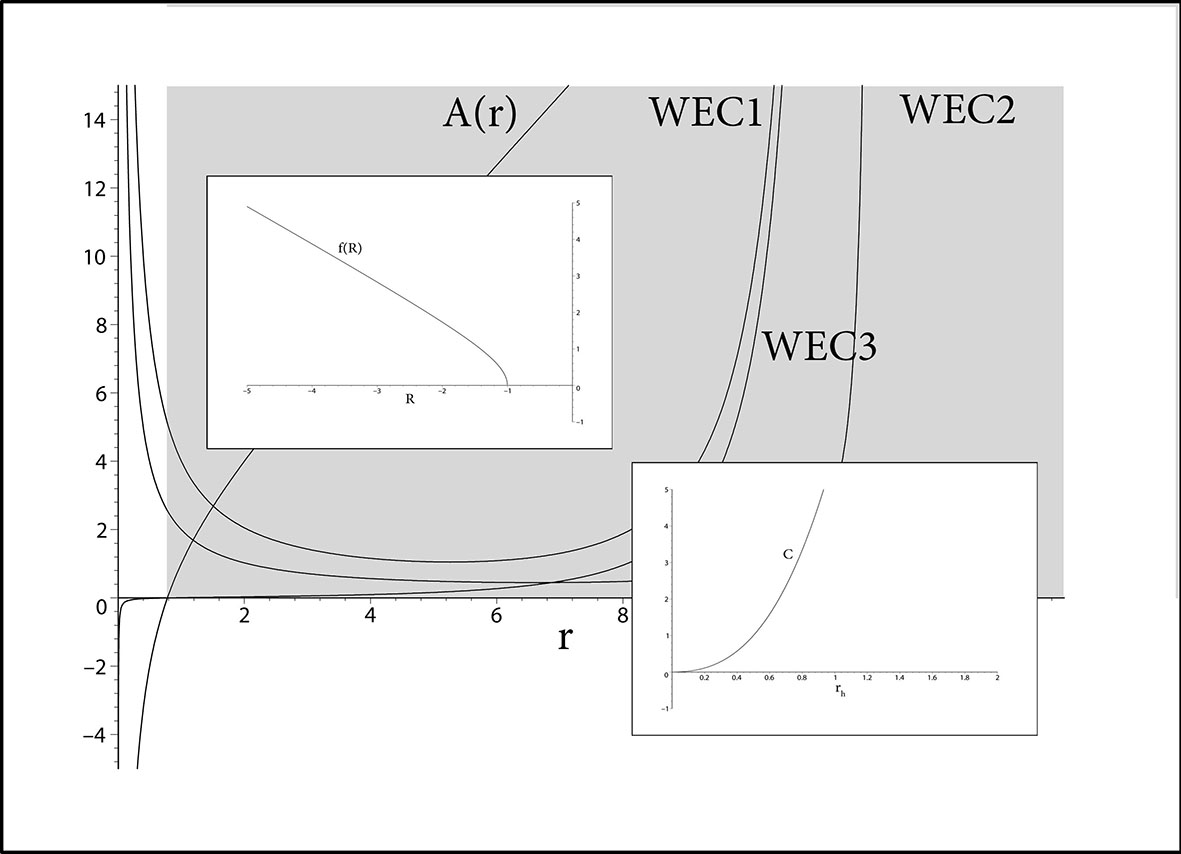}
\caption{This is the model with $f\left( R\right) =\left( \mid R\mid ^{\frac{%
1}{2}}-1\right) ^{\frac{1}{2}}$ which has WECs conditions all satisfied
while the stability condition is violated. It is thermodynamically stable
since $C>0.$}
\end{figure}

In Einstein's general relativity which corresponds to $f\left( R\right) =R$
, Rindler modification of the Schwarzschild metric faces the problem that
the energy conditions are violated. For a resolution to this problem we
invoke the large class of $f\left( R\right) $ theories. From cosmological
standpoint the main reason that we insist on the Rindler acceleration term
can be justified as follows: at large distances such a term may explain the
flat rotation curves as well as the dark matter problem. Our physical source
beside the gravitational curvature is taken to be a fluid with equal angular
components. Being negative the radial pressure is repulsive in accordance\
with expectations of the dark energy. Our scan covered ten different $%
f\left( R\right) $ models and in most cases by tuning of the free parameters
we show that WECs are satisfied. All over in ten different models we
searched primarily for the validity of WECs as well as for $\frac{d^{2}f}{%
dR^{2}}>0$, i.e. the stability. With some effort thermodynamic stability can
also be checked through the specific heat. With equal ease $\frac{df}{dR}>0,$
i.e. absence of ghost can be traced. Fig. 1 for instance, depicts the model
with $f(R)=\sqrt{R^{2}+b^{2}}$, ($b=$constant) in which WECs and stability,
even the thermodynamic stability are all satisfied, however, it hosts ghosts
since $\frac{df}{dR}<0$ for $R<0$. Finally, among all models considered
herein, we note that, Fig. 7 satisfies WECs, stability conditions as well as
ghost free condition for $r>r_{\min }$ in which $r_{\min }\geq r_{h}$
depends on the other parameters.

Finally we comment that abundance of parameters in the $f\left( R\right) $
theories is one of its weak aspects. This weakness, however, may be used to
obtain various limits and for this reason particular tuning of parameters is
crucial. Our requirements have been weak energy conditions (WECs), Rindler
acceleration, stability and absence of ghosts. Naturally further
restrictions will add further constraints to dismiss some cases considered
as viable in this study.

\bigskip

\end{document}